\documentclass[aoas]{imsart}
\RequirePackage{amsthm,amsmath,amsfonts,amssymb}
\RequirePackage[authoryear]{natbib}
\RequirePackage[colorlinks,citecolor=blue,urlcolor=blue]{hyperref}
\RequirePackage{graphicx}
\usepackage[inline]{enumitem}
\usepackage{subcaption}
\newcommand{\PP}{\ensuremath{\mathbb{P}}}
\def\GPD{{\rm GPD}}
\def\T{{ \mathrm{\scriptscriptstyle T} }}
\def\expit{{\rm expit}}
\usepackage{caption}

\begin{document}

\begin{frontmatter}
\title{Bayesian modeling of insurance claims for hail damage}
\runtitle{Bayesian hail damage modeling}

\begin{aug}
\author[A]{\fnms{Oph\'elia}~\snm{Miralles}\ead[label=e1]{ophelia.miralles@epfl.ch}},
\author[A]{\fnms{Anthony C.}~\snm{Davison}\ead[label=e2]{anthony.davison@epfl.ch}},
\author[B,C]{\fnms{Timo}~\snm{Schmid}\ead[label=e3]{timo.schmid@usys.ethz.ch}},
\address[A]{Institute of Mathematics, \'Ecole Polytechnique F\'ed\'erale de Lausanne (EPFL), Switzerland}
\address[B]{Institute for Environmental Decisions, ETH Zurich, Switzerland}
\address[C]{Federal Office of Meteorology and Climatology MeteoSwiss, Zurich, Switzerland \\\printead[presep={ }]{e1}\printead[presep={,\ }]{e2}\printead[presep={,\ }]{e3}}
\end{aug}

\begin{abstract}
Despite its importance for insurance, there is almost no literature on statistical hail damage modeling. Statistical models for hailstorms exist, though they are generally not open-source, but no study appears to have developed a stochastic hail impact function. In this paper, we use hail-related insurance claim data to build a Gaussian line process with extreme marks to model both the geographical footprint of a hailstorm and the damage to buildings that hailstones can cause. We build a model for the claim counts and claim values, and compare it to the use of a benchmark deterministic hail impact function. Our model proves to be better than the benchmark at capturing hail spatial patterns and allows for localized and extreme damage, which is seen in the insurance data. The evaluation of both the claim counts and value predictions shows that performance is improved compared to the benchmark, especially for extreme damage. Our model appears to be the first to provide realistic estimates for hail damage to individual buildings. 
\end{abstract}


\end{frontmatter}

\section{Introduction}

Global warming has already begun to affect the behaviour of insurers worldwide, both by increasing premiums and by making companies unwilling to underwrite some risks; a recent example is the May 2023 decision by the US company State Farm to cease offering house insurance to new clients in California. %The greater vigour of storms due to increases in the energy in the atmosphere is likely to lead to increased damage in the decades to come.   
Hail is of particular interest to Swiss insurance companies because of large annual insured losses, averaging several million Swiss francs (CHF) \citep{botzen2010climate}, though there is substantial year-to-year variability. Northern Switzerland experienced significant hail events in 2021, leading to estimated insured losses of 2 billion CHF and placing a heavy financial burden on insurers \citep{muller2021}. The risk of large losses due to hailstorms is increasing as a result of the construction of more new buildings each year, and climate change may increase the frequency of damaging hailstorms in Europe more broadly \citep{Raedler2019}. Destructive hailstorms are also an important risk for agriculture, buildings, and vehicles elsewhere in the world \citep{stanley2008temporal, warren2020radar}, but despite their importance, such storms can be very localised and are hard to model. 

The literature on the statistical modeling of the impact of hailstorms on buildings is very limited. Although stochastic models for hailstorm risk \citep{deepen2006schadenmodellierung, otto2009modellierung, punge2014new, pucik2017future} or hailstone size \citep{perera2018probabilistic, liu2021statistical} exist, most open-source studies on hailstorm impact use deterministic functions to link the intensity of a hail hazard to its monetary damage. The spatial footprint of hail events has been discussed in a few recent modeling studies, in which hailstorms are either represented as ellipses \citep{otto2009modellierung} or stretches of constant width \citep{deepen2006schadenmodellierung}. \citet{punge2014new} use a Poisson distribution to estimate the frequency of hail in Europe on a $50\times30$km grid, using a bimodal normal distribution in each grid cell to estimate the pointwise probability of hail based on hail reports and observations of overshooting cloud tops, whose presence for over ten minutes can indicate thunderstorm severity. \citet{deepen2006schadenmodellierung} generates a stochastic catalog of hail events in Germany by simulating areas of fixed width, random location, and random length. Hailstorms are also simulated by modeling random hailstones in \citet{liu2021statistical}, while \citet{perera2018probabilistic} propose a hailstone size distribution to aid in impact estimation. 

On the damage modeling side, \citet{hohl2002hailfall} propose a logistic impact function derived from hail kinetic energy, and \citet{schuster2006relationship} further explore the link between this energy and impacts. Claim data from insurance companies, though not usually publicly accessible, is a valuable source of information on hail impacts and has been used to complete the radar signal for hail in several recent studies. The hail damage model developed in \citet{schmidberger2018hail}, for example,  derives hail tracks in Germany from radar and insurance data, and \citet{deepen2006schadenmodellierung} uses simulated hail footprints to model damage to cars through a Poisson distribution fitted with vehicle insurance data. \citet{brown2015evaluating} use insurance data to explore the link between roof material and hail impact on buildings in Texas. 

These studies all involve randomness from the hail event itself. Indeed, radar-based proxies are used to derive the probability and/or the expected intensity of a hailstorm on grids with resolution of several kilometers. Those proxies are often chosen over direct hail measurements with automatic hail sensors \citep{kopp2023summer} because they are more spatially consistent. However, the monetary impacts due to hailstorms appear in very narrow and localized tracks that are usually poorly represented by models with such grids. 

The goal of the present study is to propose a spatially consistent model for insurance claims related to hail damage at the building level. This model differs from previous ones, as the probability and intensity of a hail event are supposed to be known, and stochasticity comes from the possible spatial impacts of a hail storm. The model we develop seems to be the first to combine a random line process and an extreme value model in order to represent hail damage tracks accurately. We describe the data available to us in Section~\ref{sec:data}, introduce a line model with extreme marks in Section~\ref{sec:model}, and describe the results we obtain when applying this model in the Swiss canton of Z\"urich, henceforth ``the canton'', in Section~\ref{sec:results}.

\section{Data and initial analysis}
\label{sec:data}
\subsection{Data}
\begin{figure}[!t]
    \centering
    \includegraphics[width=\linewidth]{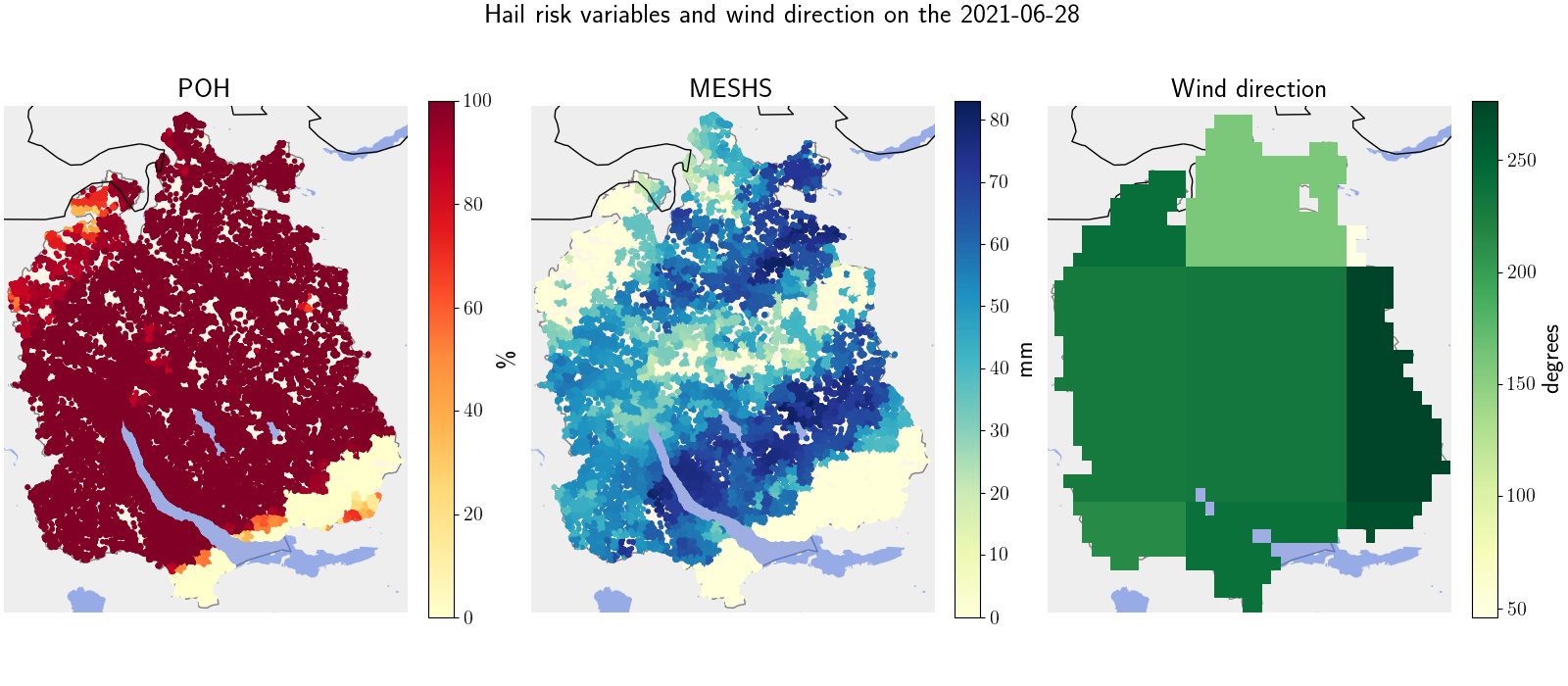}
    \caption{Hail risk covariates POH (left), MESHS (center), and wind direction (right) on 28 June 2021 in the canton ($47.15$--$47.70^{\circ}$N, $8.35$--$8.99^{\circ}$E).}
    \label{fig:exploratory_climate_variables}
    \end{figure}

Two variables representing hail risk were provided by the meteorological service MeteoSwiss in the scope of the \href{https://scclim.ethz.ch/}{scClim project}, the purpose of which is to combine knowledge from different fields to create a continuous model chain from simulating thunderstorms to quantifying the monetary impacts of hail in Switzerland. Gridded one-kilometer-resolution maps of the probability of hail (POH) and the maximum expected severe hail size (MESHS), derived from volumetric radar reflectivity \citep{nisi2016spatial}, are available for the canton during the convective season (April--September) for the years 2002--2021. The MESHS offers more spatial granularity than the POH and thus was preferred (Figure~\ref{fig:exploratory_climate_variables}).

The output from a MESHS-based deterministic damage function developed during the scClim project \citep{timo_schmid_open-source_2023} and calibrated with insurance data in the canton is also available for the same period; the results in that paper use the same function,  calibrated over several Swiss cantons. The data made available to us contain predictions for the numbers of affected assets and the monetary hail damage in each cell of a 2km square grid. For conciseness below we shall use the terms ``grid cell'' or sometimes just ``cell'' in reference to this grid. The PAA and damage functions were developed following the hazard/exposure/vulnerability methodology of the CLIMADA framework described by \citet{climada2019aznar} and will be simply referred to as ``CLIMADA'' below. We use a per-building version of the CLIMADA output, referred to as ``downscaled CLIMADA'' below, that will be used as input for our claim value model. This per-building damage is a naive downscaling of the per-cell CLIMADA damage, and attributes weights to each building as a function of its insured value. In practice, this means that every building in the cell is impacted when CLIMADA predicts a positive per-cell value, artificially inflating the number of buildings affected. 

In addition to hail-related variables, we use wind direction from the state-of-the-art ERA5 reanalysis from the European Centre for Medium-Range Weather Forecasting \citep{hersbach2020era5}, available from 1979 onwards on a 25km square grid over Europe.

\begin{figure}[!t]
    \centering
    \includegraphics[width=.8\linewidth]{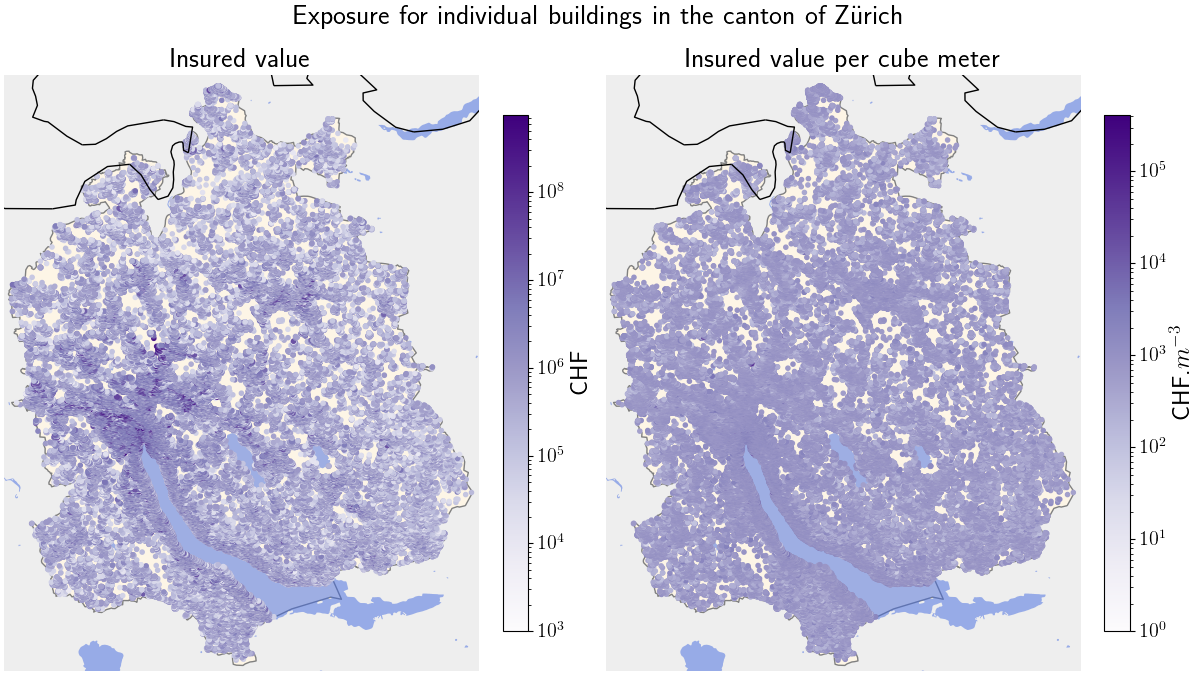}
    \caption{Insured monetary value for individual buildings (left) and the exposed monetary value per $m^{3}$ (right).}
    \label{fig:exposure}
    \end{figure}

Insurance data for hail-related claim damage to buildings in the canton is also available from one of the stakeholders of scClim, the Z\"urich cantonal insurance company GVZ. These data consist of individual claim values for buildings for the period 2000--2022, during which there were 244 days with positive claims somewhere in the canton and a total of 46254 claims. These are the amount finally paid by the insurance company in Swiss francs (CHF) following hail damage to a building and not estimated values of monetary damage. The construction year, volume, and actualized insured value of every insured building in the canton are also available.
We did not explore potential issues linked with preferential sampling, since the owner of every building in the canton is legally obliged to take out natural hazards insurance with GVZ. Consequently, data are very dense, though there are spatial disparities in exposure owing to variations in population density. As exposure equals the insured monetary value of buildings, urban areas are much more exposed than suburban areas or the countryside, but the distribution of buildings is spatially rather homogeneous in terms of average exposed value per cubic meter; see Figure~\ref{fig:exposure}. 
Our claim value model should be able to predict the difference between the true damage, i.e., the claim values reported by GVZ, and downscaled CLIMADA damage when a positive claim was recorded by GVZ. We call this target variable the ``residual damage''.

\subsection{Exploratory analysis}
\subsubsection{Hail footprints}
\label{sec:footprints}
\begin{figure}[!t]
    \centering
    \includegraphics[width=\linewidth]{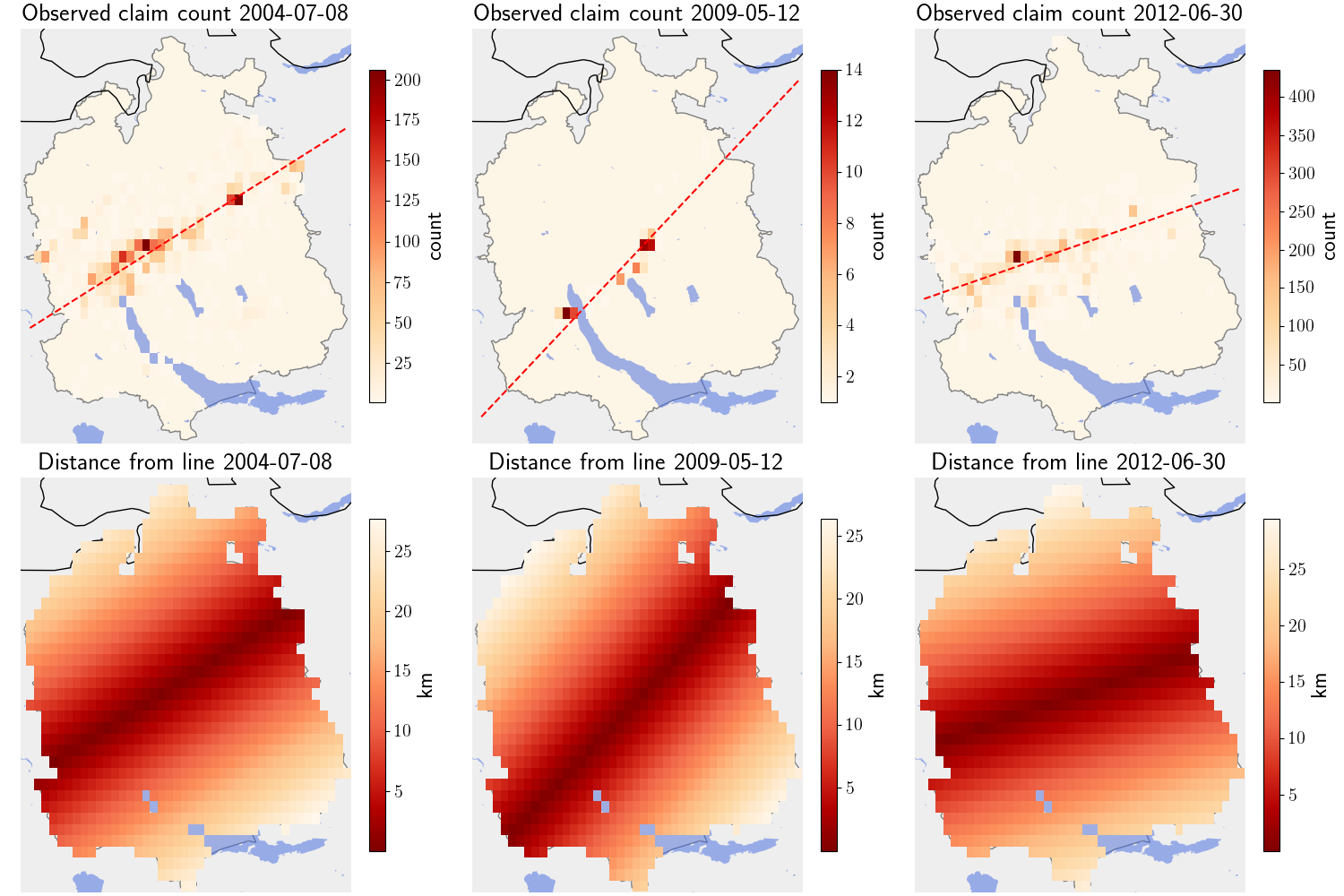}
    \caption{Example days with more than 50 recorded claims. Red-colored squares in the first row correspond to the aggregated claim count per 2km grid cell. The red line has a slope corresponding to the average wind direction on that day. The graphs on the second row represent the distance from the centroid of a 2km grid cell to the red line.}
    \label{fig:exploratory_lines}
    \end{figure}

 Literature exploring hailstorm patterns agrees on an ellipsoidal shape \citep{otto2009modellierung, punge2014new}, or a sufficiently wide straight sketch \citep{deepen2006schadenmodellierung} for modeling the spatial extent of a hail footprint. Two of those studies explore hail tracks in Germany \citep{deepen2006schadenmodellierung, otto2009modellierung} while the third treats hailstorm footprints in Central Europe \citep{punge2014new}. Although modeling single hail events on a large territory with locally bounded shapes seems reasonable, the canton of Z\"urich covers a much smaller area than Germany or Central Europe, and individual hail-related claims in our data suggest that hail storms progress along a line in space with a very narrow lateral dispersion; see Figure~\ref{fig:exploratory_lines}. Exploratory analysis suggested that the line direction is related to the average wind direction on the day of a hail storm. 

\subsubsection{Local and extreme damage}
\label{sec:extremes}
\begin{figure}[!t]
    \centering
    \includegraphics[width=\linewidth]{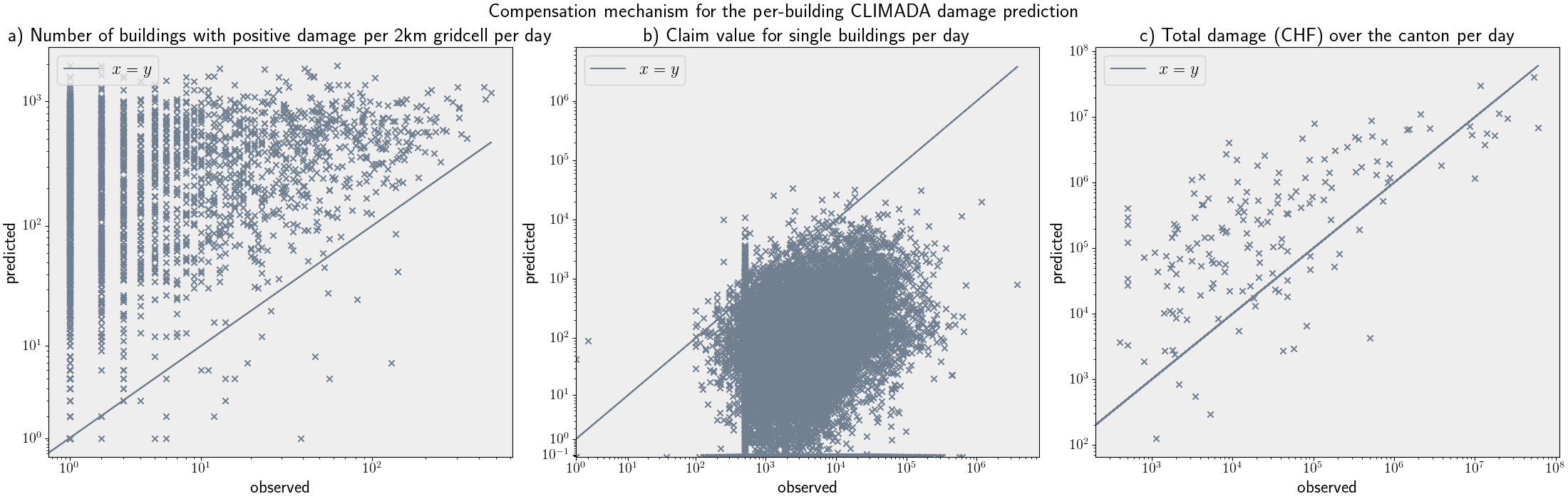}
    \caption{Number of buildings with positive predicted damage per grid cell per day using the downscaled CLIMADA impact function: (a) claim value per single building per day: (b) and total damage recorded on a given day aggregated over the canton: (c) all with the line $x=y$.}
    \label{fig:compensation}
    \end{figure}

The deterministic damage function developed through CLIMADA provides good estimates for the amount of monetary damage on a grid and adequately represents spatial patterns over the canton. However, the naively downscaled CLIMADA compensates mispredicted individual claim damage values with a very large number of claims (see Figure~\ref{fig:compensation}) as damage is distributed over all exposed buildings within a cell. Indeed, CLIMADA cannot distinguish between a few claims of high damage and many claims with low damage in a cell. Furthermore, increases in the frequency or intensity of hail would impact either the count or the value of hail-related damage, causing the compensation mechanism described above to fail. There is thus a need for a model to provide realistic values of the damage per building, which is highly relevant for insurance. The objective of this study is to provide such a model for hail-related monetary damage that respects the count/size ratio observed in the claim dat, by lowering the frequency of positive claims while allowing their values to be locally extreme. 

\begin{figure}[!t]
    \centering
    \includegraphics[width=\linewidth]{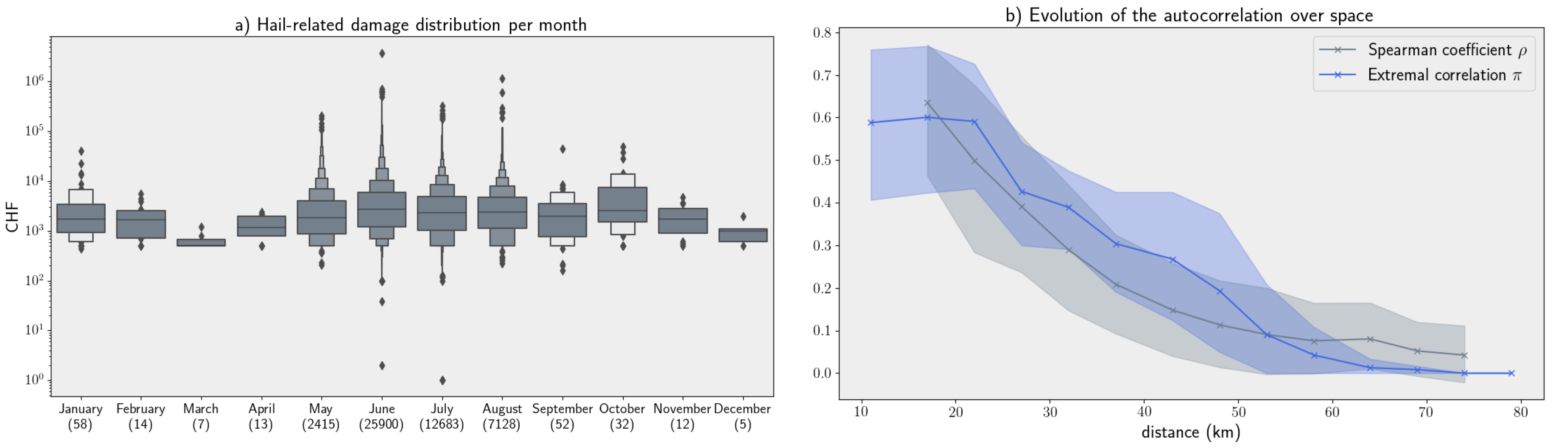}
    \caption{Exploratory analysis of claim values:  (a) average total claim value over the canton per month: (b) Spearman correlation $\rho$ and extremal correlation $\pi$ for pairwise time series of claim values per cell as a function of the distance between cells. In (b) the solid lines represent the average pairwise correlation over equally distant cells, and the shaded areas show the $90\%$ confidence range.}
    \label{fig:exploratory}
    \end{figure}

Figure~\ref{fig:exploratory} shows strong seasonal variation in claims: most damage occurs between June and August, with a peak in June, and claims occurring in April or September look less heavy-tailed than in May--August. Henceforth we only consider claims in April--September, which represent 99.74\% of the total data, and define the ``hail season'' to be May--August.

\subsection{Spatiotemporal correlation}
\label{sec:corr}
Claims for damage from a hail event can be made on that day or with a lag of a few days, so the damage function derived through CLIMADA pre-processes the original insurance data to cluster claims received during a 4-day window around a big hail event detected with the POH values. For any reported claim, POH values in a $\pm 2$-day window are scanned. If a POH higher than 50\% of that on the day of the reported claim is observed, the claim date is changed to the day with the highest POH \citep{timo_schmid_open-source_2023}.
This pre-processing step largely succeeds in removing short-term autocorrelation in the claim values, so here we focus on spatial correlation. For this purpose, we introduce the extremal correlation $\pi_h(u) = \PP(X_{s+h} \geq u \mid X_s \geq u)$ of the variable $X_s$, where $s$ denotes the spatial location of a cell, $u$ is a high threshold and $h$ a spatial lag. The threshold $u$ is chosen by applying the threshold selection method described in \citet{varty2021inference} to the log total sum of damage $X_s$ (see Supplementary Material). We also study the pairwise Spearman correlation $\rho$ for the daily sum of claim values per cell.
Figure~\ref{fig:exploratory} shows that both $\pi$ and $\rho$ decrease as the distance between two spatial locations increases.

\section{Key model elements}
\label{sec:model}
In this section we describe the key elements of a model to reproduce the very localized but large damage seen in the data. We first explain how the long and narrow hail footprint (as observed in Section~\ref{sec:footprints}) can be modeled using a Gaussian line process, and then recall the peaks-over-threshold approach from extreme value theory, which is used to account for large impacts on individual buildings (see Section~\ref{sec:extremes}).  More details of the modeling are given in Section~\ref{sec:twostepmodel}. 

\subsection{Random line process}
\label{sec:lines}

Claim values are usually represented as a spatiotemporal point process $s_t = (t, x, y)$, with $t\geq0$ and $(x, y)$ the geographical coordinates. In the forthcoming discussion, a sequence of distance-conserving transformations will be applied to map this onto a coordinate system that is better suited for defining the random line model.

For $t>0$ let $\Theta_t\in [-\pi, \pi]$ and $\alpha_t$ respectively be the time-varying random angle and vertical deviation of a line $\mathcal L$ from a chosen origin point $s_0=(x_0, y_0)$,  defined as the set of points
\begin{equation}
    \mathcal{L} = \left\{(s, t) : s^\T \begin{pmatrix}
        -\tan{\Theta_t} \\
        1
    \end{pmatrix} =  \alpha_t \right\},
    \label{eq:line_process}
\end{equation}
where $s=(x, y)$ denotes geographical coordinates and $t$ is the discrete time coordinate. At time $t$ the projection of any pair of spatial coordinates $s=(x, y)$ in the coordinate system in which $\mathcal{L}$ is the horizontal axis and the origin is $s_0$ may be written as
\begin{equation}
        s_t^{\mathcal{L}}= \begin{pmatrix}
    \cos{\Theta_t} & \sin{\Theta_t}\\
    -\sin{\Theta_t} & \cos{\Theta_t}
    \end{pmatrix} \begin{pmatrix}
    x \\ y-\alpha_t
    \end{pmatrix},
\end{equation}
and the orthogonal projection of $s$ onto the line would thus be the point $\pi_t^{\mathcal{L}}(s)=(\cos{\Theta_t x} + \sin{\Theta_t}y, 0)$ in the new coordinate system. The Euclidean distance between a point in space and the line $\mathcal{L}$ can be computed in any coordinate system, and as $s_t^{\mathcal{L}}$ and $\pi_t^{\mathcal{L}}(s)$ have the same  $x$-coordinate in the new system, at time $t$  this distance  can be expressed as
\begin{equation}
d_t\left(s, \mathcal{L}\right) = \vert y'\vert = \vert (y-\alpha_t)\cos{\Theta_t} - x\sin{\Theta_t} \vert.
    \label{eq:distance}
\end{equation}

To allow the intensity of points at time $t$ to be highest close to the random line $\mathcal{L}$, we define a spatiotemporal Gaussian field $X^{\mu}(s, t)$ whose mean is 
\begin{equation}
    m_t(s) = \dfrac{\sigma_m}{1+d_t\left(s, \mathcal{L}\right) }-1,
    \label{eq:mt}
\end{equation}
where $\sigma_m$ is a dispersion parameter that controls the concentration of points around $\mathcal{L}$. In accordance with the exploratory analysis, we chose a correlation function $\rho$ such that the correlation between $X^{\mu}(s_0, t)$ and $X^{\mu}(s_1, t)$ decreases when the distance $w$ between $s_0$ and $s_1$ increases (see Figure~\ref{fig:exploratory}). A common choice is the Mat\'ern correlation function
\begin{equation}
    \rho(w) = \{2^{\nu-1} \Gamma(\nu)\}^{-1}(w/l)^\nu K_{\nu}(w/l),\quad w>0,
    \label{eq:matern_comp}
\end{equation}
where $\nu>0$ is a shape parameter controlling the smoothness of the Gaussian process, $l>0$ is a scale parameter, $\Gamma(\cdot)$ is the Gamma function, and $K_{\nu}(\cdot)$ is the modified Bessel function of the second kind. After some experimentation, we took $\nu=1.5$, which gives fields of similar smoothness to the data, and estimate the parameter $l$ as part of a hierarchical Bayesian model. When $\nu=1.5$, Equation~\eqref{eq:matern_comp} simplifies to
\begin{equation}
    \rho(w) = \left(1+{\sqrt{3}w}/{l}\right)\exp{\left(-{\sqrt{3}w}/{l}\right)}, \quad w>0.
    \label{eq:matern}
\end{equation}

\subsection{Marginal model for extreme claim values}
 The exploratory analysis suggests using extreme value theory to model the largest claim values. The generalized Pareto distribution, 
\begin{align}
    \GPD^u(x) &= 1- \left(1 + \xi \dfrac{x-u}{\sigma_u}\right)_+^{-1/\xi}, \quad x \in [u, \infty), \label{eq:gpd}
\end{align}
where $a_+=\max(a,0)$ for real numbers $a$, provides a standard model for the exceedances of a high threshold $u$. The model depends on a shape parameter $\xi$ that determines the weight of the distribution tails and on a scale parameter $\sigma_u$; both are specified in Section~\ref{sec:gpd_model}. We select a constant threshold $u$ by applying the method described by \citet{varty2021inference} to the log of the total sum of claim values over the canton; see the Supplementary Material. 

\section{Modeling extreme hailstorms}
\label{sec:twostepmodel}
Our model for hail damage uses a discrete zero-inflated count process for the number of claims and a continuous two-part distribution for hail damage values. In the following section, the spatiotemporal matrix of observed covariates will be designated by the letter $M$. If a variable needs to be specified, it is written $M^{\text{NAME}}$ --- for the MESHS, for instance, we write $M^{\text{MESHS}}$. Building exposure, MESHS, POH, CLIMADA predicted claim count and downscaled CLIMADA value are respectively designated by Exp, MESHS, POH, NC, and YC. The count of individual claims in a grid cell on a given day is denoted by $N$, while the value of an individual claim is designated by $Y$. 

\subsection{Hail damage count}
\label{sec:count_model}
The claim count is modeled on a 2km square grid.
Hail can either strike very locally and violently or can be spread out more smoothly, as observed in the data, in which the maximum number of observed claims in a cell on a single day is 469, and the nonzero minimum is 1. In view of this wide range of values, we model $N$ as a negative binomial. Positive counts are scarce, so we model them as realisations of a zero-inflated negative binomial random variable, with probability mass function
\begin{equation}
\text{NB}_{\psi, \mu, \alpha}(x) = 
\begin{cases}
           (1-\psi) + \psi \left (
             \dfrac{\alpha}{\alpha+\mu}
           \right) ^\alpha, & x = 0, \\
           \psi \dfrac{\Gamma(x+\alpha)}{x! \Gamma(\alpha)} \left (
             \dfrac{\alpha}{\mu+\alpha}
           \right)^\alpha \left(
             \dfrac{\mu}{\mu+\alpha}
           \right)^x, & x=1,2,3,\ldots, 
         \end{cases}
    \label{eq:counts}
\end{equation} 
where $\psi\in(0,1)$, $\mu>0$ and $\alpha>0$ is a shape parameter. We set $N \mid \psi, \mu, \alpha \sim \text{NB}_{\psi, {\mu}, \alpha}$.

The probability $\psi$ of observing a non-zero claim in grid cell $s$ on day $t$ is modeled as 
\begin{equation}
\psi(s,t) = \expit\left\{\psi_0 + \psi_1 \mathbf{1}_{M^{\text{NC}}_{s, t}>0} + \psi_2 M^{\text{NC}}_{s, t} m_t(s)\right\},
    \label{psi_poisson}
\end{equation}
where $\expit(x) = \{1+\exp(-x)\}^{-1}$ and $\psi_0,\psi_1,\ldots$ are real parameters. We define the mean $\mu$ of the negative binomial variable through the equation
\begin{equation}
 \begin{aligned}
\log  \mu(s, t) = \mu_0 + \sum_{i=1}^3 \mu_{1i} \left(M_{s, t}^{\text{NC}}\right)^i + \mu_2M_{s, t}^{\text{NC}} m_t(s) + X^{\mu}(s,t)+ \epsilon(t),
\end{aligned}
    \label{mu_poisson}
\end{equation}
where $X^{\mu}$ is a spatiotemporal Gaussian field whose mean $m_t$ and covariance function $\rho$ are given respectively in~\eqref{eq:mt} and~\eqref{eq:matern}, and $\mu_0, \ldots $ are  real parameters. The Gaussian noise $\epsilon(t) $ has mean zero and one variance for the months April and September and another variance for the months May--August. 

\subsection{Hail damage values}
The number of positive claims (46,254) is much smaller than the roughly 350,000 cell-date combinations in which hail events might have occurred, so it is reasonable to model spatial patterns at a coarser resolution than the 2km grid used for the counts. Spatial Gaussian random fields used to model unobserved covariates underlying the hail damage values are thus defined over a grid of resolution roughly 10km in which each cell has at least 100 positive claims over the years 2000--2015. 

Downscaled CLIMADA under-predicts the values of $99.3\%$ of reported claims, so we model only positive errors, i.e., if the predicted count $N$ is positive, we only allow a shift upwards from the downscaled CLIMADA value $M^{\text{YC}}$. The resulting residual hail damage variable $Z=Y-M^{\text{YC}}$ is modelled using a beta model for non-extreme values and a generalised Pareto model for extreme values. 

We first introduce a binary variable $R$ to model the event that a claim value exceeds the threshold $u$ ($R=1$) or not ($R=0$), with success probability 
\begin{equation}
    p(s, t) = \expit\left\{p_0 + p_1M_{s, t}^{\text{POH}}+p_2M_{s, t}^{\text{MESHS}}+p_3M_{s, t}^{\text{MESHS  $\cdot$ POH}}+p_4M_{s, t}^{\text{Exp}}+\chi(s) +\epsilon_p(t)\right\},
\end{equation}
where $M_{s, t}^{\text{MESHS  $\cdot$ POH}}=M_{s, t}^{\text{MESHS}}M_{s, t}^{\text{POH}}$ and $\chi$ and $\epsilon_p$ normally-distributed random effects respectively per grid cell and season. The Beta and Pareto models for non-extreme and extreme claim values are detailed below. In the following sections, $f$ denotes the function $f: x \mapsto \log(1+x)$.

\subsubsection{Non-extreme residual damage}
\label{sec:beta_model}
Residual damage $Z$ for which $f(Z)\leq u$ is described by letting ${Z}/{f^{-1}(u)}$ have a beta density with mean $\nu$ and variance ${\nu(1-\nu)}/{\kappa+1}$,  
\begin{equation}
      \textrm{Beta}_{\nu, \kappa}(x) = \dfrac{x^{\nu\kappa-1}(1-x)^{(1-\nu)\kappa-1}}{B\left\{\nu\kappa, (1-\nu)\kappa\right\}}, \quad x \in (0, 1), 
        \label{eq:beta}
\end{equation}
where $B(\alpha, \beta) = {\Gamma(\alpha)\Gamma(\beta)}/{\Gamma(\alpha+\beta)}$.  We set 
\begin{equation}
    \dfrac{Z}{f^{-1}(u)} \quad \mid \quad \left\{f(Z) \leq u\right\}, \mu_B, \sigma_B \sim  \text{Beta}_{\nu, \kappa},
\label{eq:beta_param}
\end{equation}
and model the mean of this variable via the expression 
\begin{equation}
    \nu_t(s) = \expit\left\{\nu_0 + \nu_1M_{s, t}^{\text{POH}} + \nu_2M_{s, t}^{\text{MESHS}} + \nu_3 M_s^{\text{Exp}}+ X^{\beta}(s)\right\}, 
    \label{eq:nu}
\end{equation}
where $X^{\beta}$ is a spatial Gaussian process with zero mean and covariance kernel function
\begin{equation}
    \rho_{\alpha}(w) = \left(1+\dfrac{w}{4 l^2}\right)^{-2}.
    \label{eq:ratquad}
\end{equation}

\subsubsection{Extreme residual damage}
\label{sec:gpd_model}
Damage arising when $f(Z)>u$  is modeled by letting $f(Z)-u$ be a generalized Pareto variable, 
\begin{equation}
    f(Z)-u \,\mid\,  \{f(Z)>u\}, \sigma_u, \xi \sim  \GPD_{ \sigma_u, \xi};
\label{eq:gpd_mod}
\end{equation} 
the distribution function is given in~\eqref{eq:gpd}.  The extremal correlation observed in Figure~\ref{fig:exploratory} is accommodated by allowing $\sigma_u$ to depend on  POH, MESHS, exposure covariates and unobserved spatial discrepancies and time-related autocorrelation, respectively modeled with a Gaussian process and an auto-regressive process, leading to
\begin{equation}
       \log \sigma_{u, t}(s) = \sigma_0 + \sigma_1M_{s, t}^{\text{MESHS}} + \sigma_2M_{s, t}^{\text{MESHS  $\cdot$ POH}} + \sigma_3 M_s^{\text{Exp}}+ X^{\sigma}(s)
 \label{eq:sigma_gpd}
\end{equation}
where $X^{\sigma}$ is a spatial Gaussian process with zero mean and a Mat\'ern covariance matrix~\eqref{eq:matern}, in which the Euclidean distance has been replaced by the chordal distance because our Gaussian process occurs on the surface of a sphere, whose curvature should be reflected by our model \citep{jeong2017spherical}. The chordal distance is the length of a line passing through the three-dimensional Earth to connect two points on its surface. For two locations $s_1$ and $s_2$ with respective geographical coordinates $(x_1, y_1)$, $(x_2, y_2)$, the chordal distance between $s_1$ and $s_2$ is defined by 
\[{\rm C}(s_1, s_2) = 2r \arcsin\left[\frac12\left\{1 - \cos_p(y_2-y_1) + \cos_p y_1\cos_p y_2(1-\cos_p (x_2-x_1)\right\}\right]^{1/2}, \]
where $r=6371$km is the Earth's radius and $\cos_p(z) = \cos(\pi z/180)$. In a small area such as the canton, using the chordal distance instead of the Euclidean distance might not make a huge difference, but it would matter if the model was used for larger regions.

In view of Figure~\ref{fig:exploratory} we allowed the shape parameter to vary as 
\begin{equation}
    \xi(t) = \begin{cases}
    \xi_1 ,&t \in \{\text{May, June, July, August}\},\\
    \xi_2, &\text{ otherwise.}
    \end{cases}
\end{equation}

\section{Model fitting and validation}
\subsection{Technical challenges}
\label{sec:fitting}
Fitting the Bayesian hierarchical model described in Section~\ref{sec:twostepmodel} is challenging due to its complexity, the size of the parameter space and the large number of data points. Recent advances in spatial statistics allow better computational efficiency for Bayesian models with latent variables \citep{rue2017bayesian}. In our case, it would be desirable to use R-INLA, which has been widely and successfully used for environmental data \citep[e.g.,][]{castro2019spliced, koh2023spatiotemporal}.
%Such models attempt to account for hidden or unobserved effects in high-dimensional data, and Gaussian variables and processes can flexibly capture local correlations and uncertainty. Latent Gaussian models combine these concepts \citep{lawrence2003gaussian, rue2009approximate} and comprise a large class of statistical models. 
%The R-INLA package \citep{rue2017bayesian} estimates posterior distributions for latent Gaussian models by approximating the target with a normal distribution using Laplace approximation. The size of the latent field influences the complexity of the precision matrix computation for the Gaussian target, and \citet{rue2017bayesian} argue that assuming Markov properties for the target process can greatly reduce the computational burden. The R-INLA package has been widely used for environmental data; \citet{castro2019spliced}, for example, use it to fit a hierarchical Bayesian model involving a two-part distribution for extreme and non-extreme wind speeds observed at 260 stations in the USA.
As our work is part of a collaboration involving several subprojects mostly written in the programming language Python, the model described in Sections~\ref{sec:model} and~\ref{sec:twostepmodel} was also coded in Python so that our collaborators would find it accessible. There is no equivalent of R-INLA  in Python, and reproducing it for Python users would have taken far too long, so despite the resulting drop in computational efficiency we resorted to Markov chain Monte Carlo (MCMC) methods.

In contrast to Metropolis--Hastings steps, which make trajectory proposals within a possibly skewed ball \citep{hastings1970monte, metropolis2004equation}, or to Gibbs sampling, which generally only moves in a few dimensions at a time \citep{gelfand2000gibbs}, Hamiltonian Monte Carlo (HMC) generates proposals based on the shape of the posterior by using its gradient \citep{betancourt2018conceptual}. In MCMC algorithms, the termination criterion identifies when a trajectory is long enough for adequate exploration of the neighborhood around the current state, but in HMC, this criterion should be chosen to compromise between taking full advantage of the Hamiltonian trajectories and wise use of computational resources \citep{betancourt2018conceptual}.
The No-U-Turn Sampler (NUTS) is a HMC algorithm that proves particularly efficient in converging for high-dimensional posterior distributions \citep{hoffman2014no}. Indeed, NUTS uses a dynamic termination criterion that considers only the position and momentum of a trajectory's boundaries: when it is met, further sampling typically leads to neighborhoods that have already been explored. In addition to this specific termination criterion, NUTS implements a multiplicative expansion of the trajectory that allows fast exploration of the parameter space within limited computer memory \citep{betancourt2018conceptual}. We used NUTS for the count model described in Section~\ref{sec:count_model} and for the extremal model described in Section~\ref{sec:gpd_model}. 

For the non-extreme claims model detailed in Section~\ref{sec:beta_model}, a differential evolution Metropolis (DE-MC) sampling step with a snooker updater was used, as it is more efficient and faster than the classical random walk Metropolis step.  DE-MC combines a differential evolution genetic algorithm and MCMC simulation \citep{ter2006markov}. The snooker updater makes it less computationally expensive than classical DE-MC, as it updates different chains in parallel with information from past states \citep{ter2008differential}, and is faster than NUTS for this model, with no significant impact on the results.

We systematically exclude the initial 500 samples drawn, which are reserved for a tuning phase during which the sampler dynamically adjusts the step sizes and scalings to optimize its subsequent performance. We monitor the convergence of the model parameters using informal diagnostic plots; see the Supplementary Material. We check that autocorrelation has decreased to approximately zero during the sampling, and examine trace plots of the sampled parameters for the absence of patterns. Running the claim counts model took about five hours for about a thousand parameters. For the claim values, fitting the model took two hours for the GPD model (38 parameters) and less than an hour for the Beta model (36 parameters). Non-informative priors were found to perform significantly better than weakly informative priors in our case and thus were attributed to all of the model parameters. To make sure the posterior is proper, we check that its distribution percentiles and mean are finite and reasonable.

\subsection{Metrics}
\label{sec:metrics}
To assess the model's performance in improving spatial patterns we use diagnostic quantities that include the following two specific metrics.

The spatially convolved Kolmogorov--Smirnov statistic \citep{miralles2022downscaling} represents the disagreement between the spatial distributions of the generated and observed images and  is computed as the maximum absolute difference of empirical cumulative distribution functions for the generated and true damage, summed over $10\times10$ patches of the image of interest. The aim is to obtain a metric with properties close to those of the Fr\'echet inception distance \citep{heusel2017gans} for images by assessing the match between predictions and targets, as a human eye would. After extracting $M$ spatial patches of constant size from the target and predicted images, we set
\[\mathrm{SKSS} =  \sum_{t \leq N_{T}, j \leq M}\max_{x\in\mathbb{R}} \vert F_{jt}(x)-\hat{F}_{jt}(x) \vert,\]
where $F_{jt}$ represents the empirical cumulative distribution function of the hail damage for a single spatial patch $j$ and time $t$ and $\hat{F}_{jt}$ is its analog for predicted damage. This metric evaluates the local agreement between two distributions rather than focusing on individual pixels.

The log-spectral distance  \citep{rabiner1993fundamentals} is expressed as the log-difference of power spectra between the generated and realized samples, 
\[\mathrm{LSD} = \left\{ \dfrac{1}{2N_{T}\times P} \sum_{t \leq N_{T}, i \leq P} \left[10\log_{10}\left( \dfrac{\vert g(c_{it})\vert^2}{\vert g(\hat{c}_{it})\vert^2} \right)\right]^2\right\}^{1/2},\]
where $g$ is the Fourier transform, $\vert g(\cdot)\vert^2$ the power spectrum, $c$ is the target map of damage and $\hat{c}$ its estimate. This evaluates whether the generated images reproduce the spatial structures noticeable in the target images.

\section{Results}
\label{sec:results}
 There is no visible long-term trend in either the claim count or value in the insurance data; we can thus split data into sets of consecutive years. The training set is built from years up to 2015, the validation data comprises the years 2016--2017, and later years are used as the test set. In the following analysis of the results, unless specified otherwise, the average prediction for the test set over 1000 different sets of parameters sampled from the posterior distribution is used to construct graphs and maps. We recall that our objectives are to accurately capture spatial patterns for hail damage, to be able to predict localized and extreme damage and to match the distribution of the target data provided by GVZ. We shall see that the fitted random line process with extreme marks achieves this. We start by evaluating the performance of the Gaussian line process, then explain the procedure for combining counts and claim values, and finally discuss predicted claims.

\subsection{Claim counts}
\begin{figure}[!t]
    \centering
    \includegraphics[width=.5\linewidth]{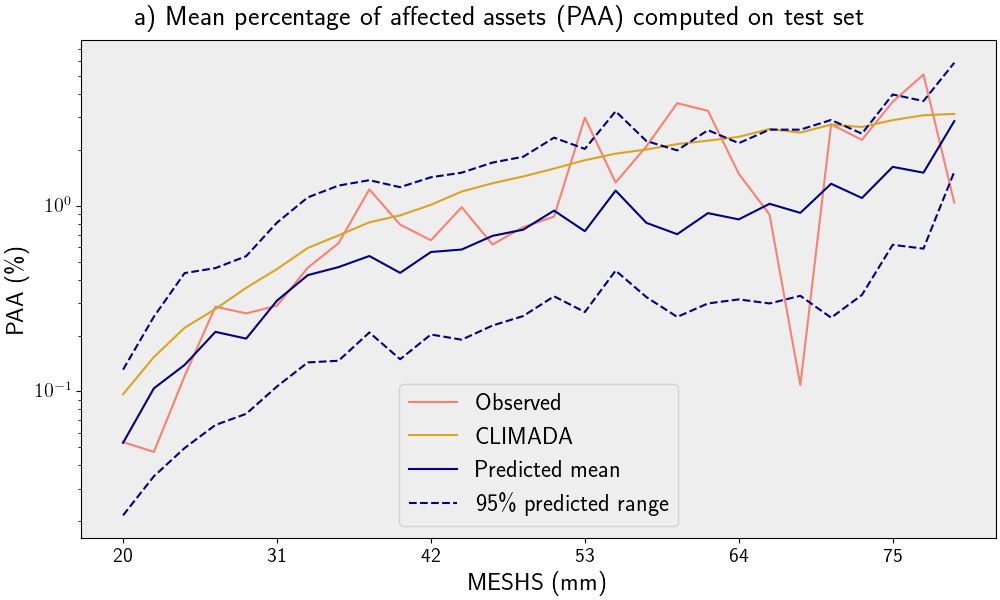}
    \includegraphics[width=.49\linewidth]{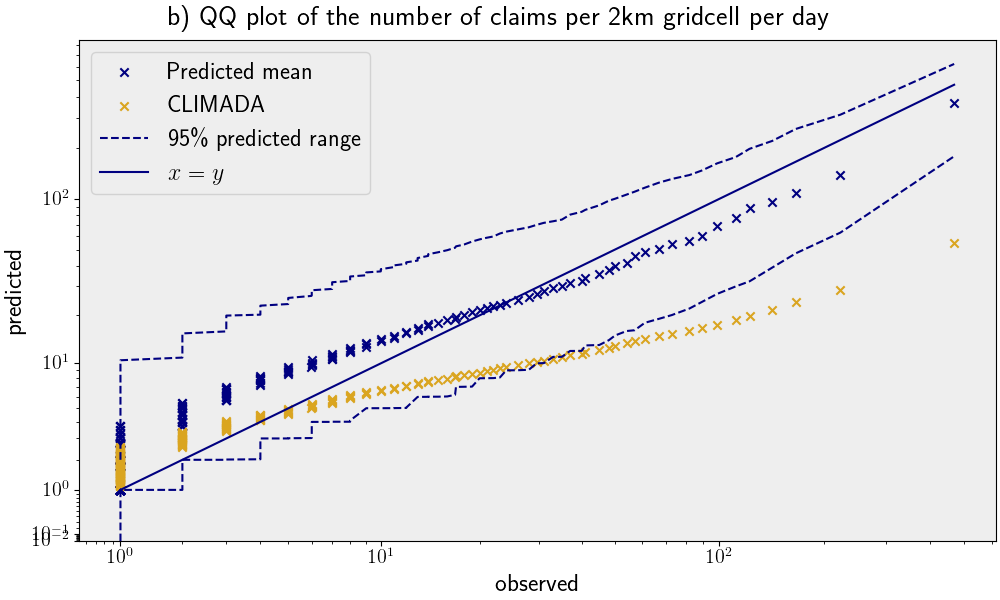}
    \caption{Comparison of CLIMADA and model:  (a) predicted and observed percentages of affected assets; (b) QQ-plots of realized versus predicted quantiles for the number of claims per grid cell per day.}
    \label{fig:counts_paa_qq}
\end{figure}

Our benchmark for evaluating the performance of the random line model presented in Section~\ref{sec:lines} is the percentage of affected assets (PAA)-based gridded claim count predicted with CLIMADA \citep{timo_schmid_open-source_2023}. The PAA, defined as the per-cell proportion of damaged buildings, is expected to increase with the value of MESHS, since hailstorms with larger hailstones should cause more damage. Figure~\ref{fig:counts_paa_qq}(a) shows that the observed PAA does not increase linearly and is very variable, but that the predicted and observed values are quite similar, while the 95\% prediction range captures the observed variation well. The impact function computed through CLIMADA tends to over-predict the percentage of affected assets for any observed MESHS value. 

Figure~\ref{fig:counts_paa_qq}(b) suggests that small claim counts are over-predicted by our model. For days with more than 1000 recorded claims, the distribution of predicted counts is very close to the observed claim counts; the line process captures days with very many claims particularly well. 

Table~\ref{btoa_table} assesses how much our model improves on CLIMADA in terms of predicting the daily claim count. The random line model reduces the false alarm rate by about 40\% and increases the positive predictive value by 15\% and the specificity by 42\%, so it makes fewer mistakes on average in predicting both positive and zero counts. Compared to CLIMADA total predicted counts per day, Table~\ref{btoa_table} shows that the sensitivity has dropped by 12\%, i.e., our model might miss days with a positive claim count, but inspection of the data reveals that it only misses days with fewer than ten claims and less than CHF~10K overall damage.

\begin{center}
    \begin{figure}[!th]
        \centering
                \includegraphics[width= \linewidth]{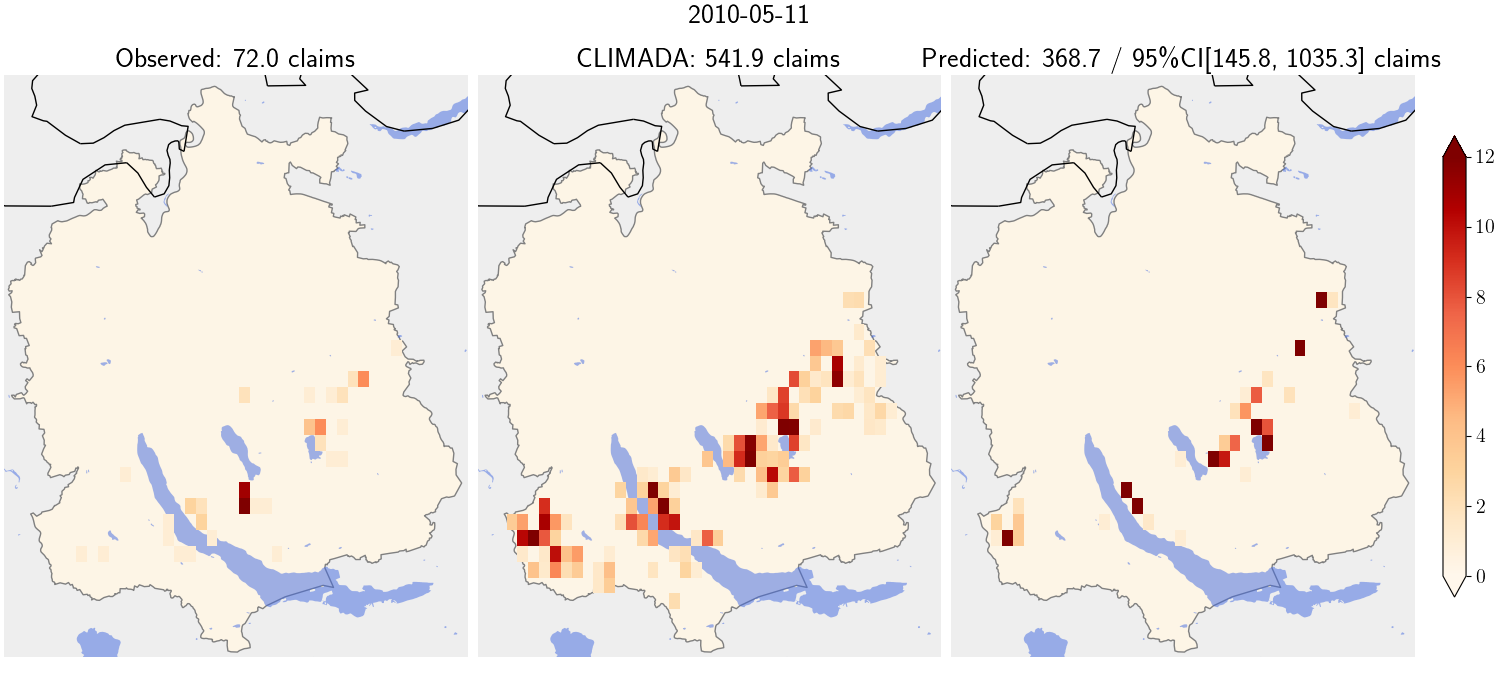}
        \includegraphics[width= \linewidth]{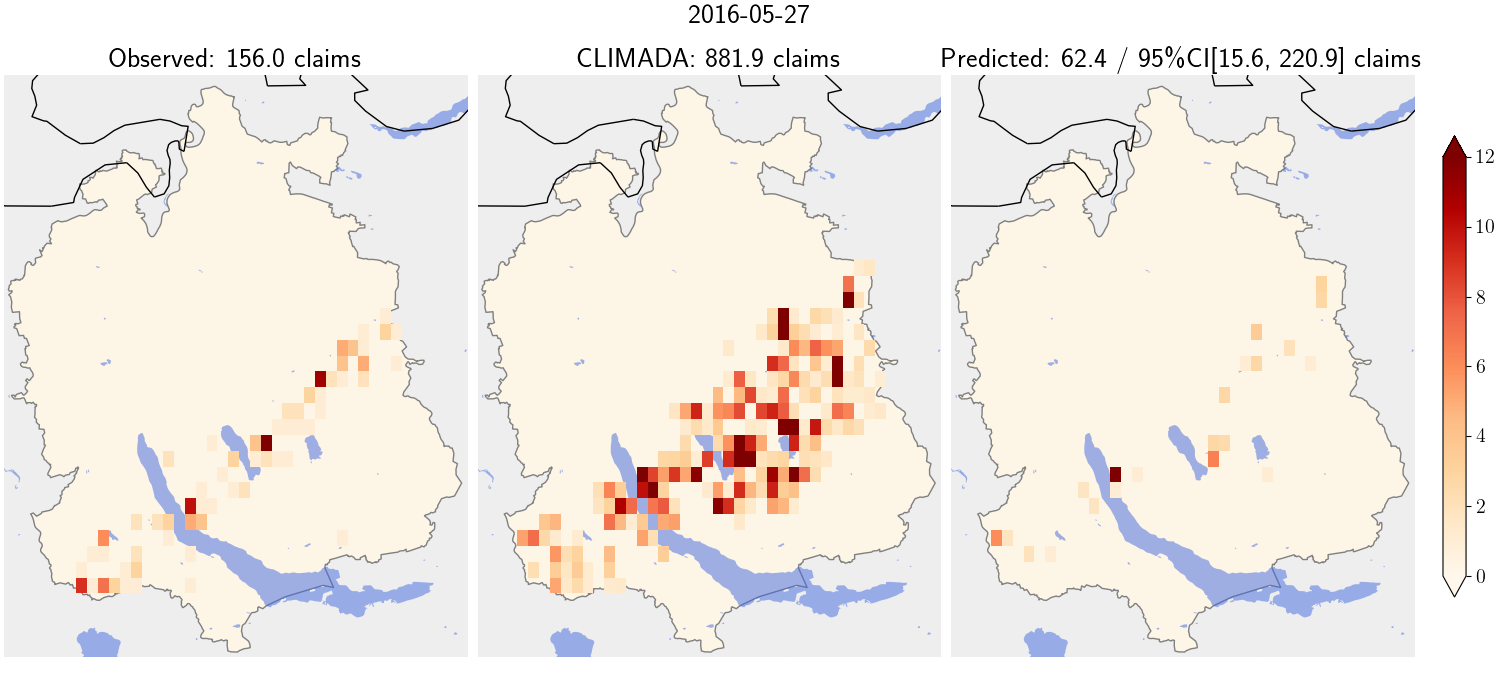}
        \includegraphics[width= \linewidth]{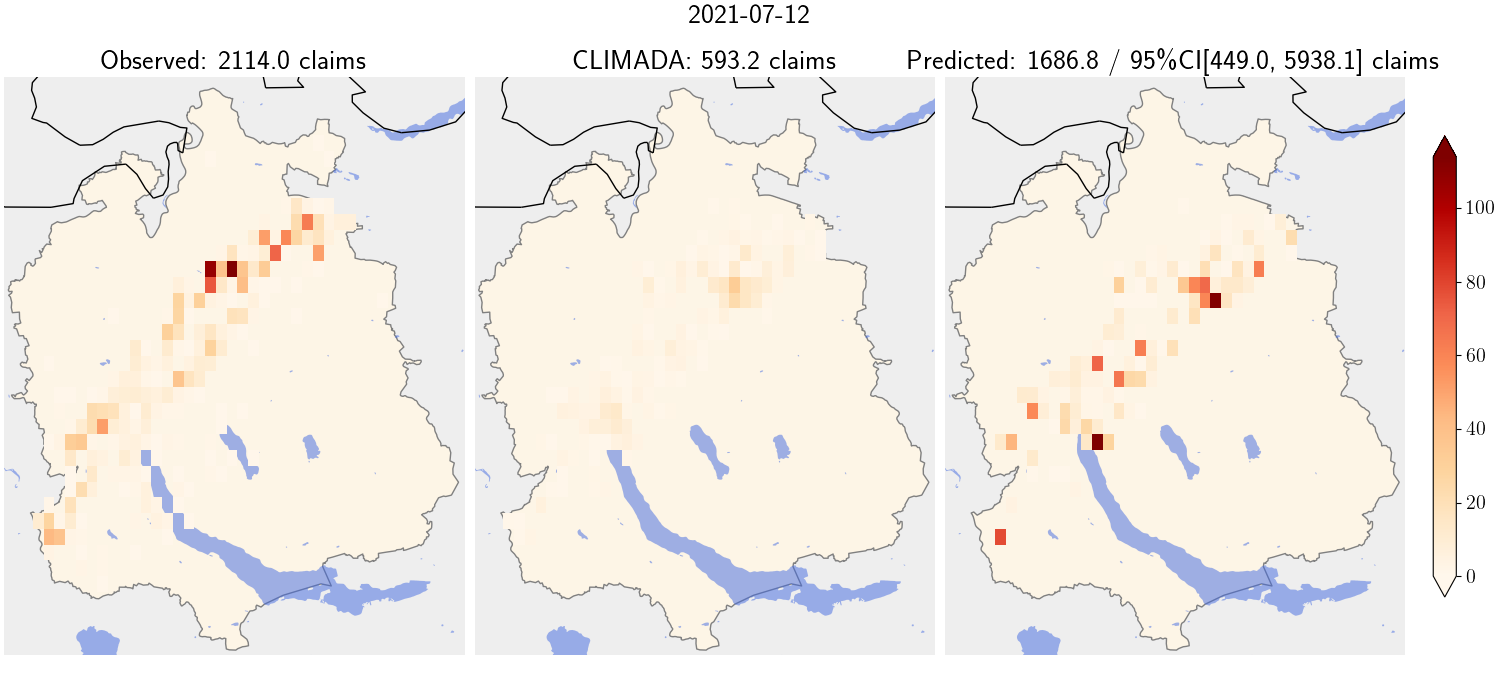}
        \caption{Comparison of locations of observed claim counts (left),  CLIMADA predicted counts (center), and our predicted counts (right) for three dates selected over all dates with more than 10 observed claims over the canton on the 2000--2021 period. The titles give the observed number of claims, CLIMADA predicted count, and the average count and its 95\% predicted range from the random line model (right).}
        \label{example_maps_counts} 
    \end{figure}
\end{center}

Examples of predicted counts plotted in Figure~\ref{example_maps_counts} show that the line model helps to concentrate the predicted damage on straight lines, giving results that resemble observed claim counts which are usually concentrated in hail streaks of width just a few km in the Alpine region \citep{nisi2018alpine}. In contrast, the predictions from CLIMADA are broadly distributed according to the MESHS footprint, which typically covers a whole storm cell core \citep{nisi2016spatial}. The average predicted count over the canton is also closer to the realized value using the line model than with CLIMADA.

\begin{table}[!t]
    \centering
    \caption{Comparison of the false alarm rate, sensitivity, specificity and positive predictive value (\%) averaged over the time dimension for CLIMADA's hail damage model and the random line model. If $a$, $b$, $c$ and $d$ denote the true positive, false positive, false negative and true negative numbers, the false alarm rate is computed as $b/(b+d)$, the sensitivity as $a/(a+c)$, the specificity as $d/(b+d)$ and the positive predictive value as $a/(a+b)$.}
    \begin{tabular}{|l|c|c|c|c|}
    \hline
        \textbf{} & \textbf{False Alarm} & \textbf{Sensitivity} &\textbf{Specificity}  & \textbf{Positive Predictive Value}\\ \hline
        \textbf{CLIMADA} & $72.1$ & $64.8$ & $27.9$ & $62.9$ \\ \hline
        \textbf{Model} &  $29.7$ & $52.1$ & $70.3$ & $77.0$ \\ \hline
    \end{tabular}
    \label{btoa_table}
\end{table}

\subsection{Combination of claim counts and values}
\label{sec:combination}

One way to combine counts and values would be to compute a per-cell impact coefficient corresponding to the predicted number of buildings impacted by a hailstorm divided by the total number of buildings in the grid cell. This would then be multiplied by the total possible hail damage in the cell, i.e., the sum of predicted values of claims for all buildings, to obtain the per-cell effective damage. This possibility was considered but not pursued, since our aim is to predict claim values for single buildings and not the aggregated damage in a cell. 
Our combination of counts and values thus involves choosing which buildings are impacted by hail given the predicted count in a cell. In each cell, buildings are first sorted by their exposure (i.e., insured value), and the first $N$ are selected to compute the total damage. To predict damage and its confidence range, we sample the counts for the cell $n$ times and the claim value for every building $m$ times, apply the procedure described in the previous sentence to the $mn$ samples, and finally compute the average total damage and its 95$\%$ prediction range.

\subsection{Claim values}

\begin{figure}[!th]
    \centering
    \includegraphics[width=\linewidth]{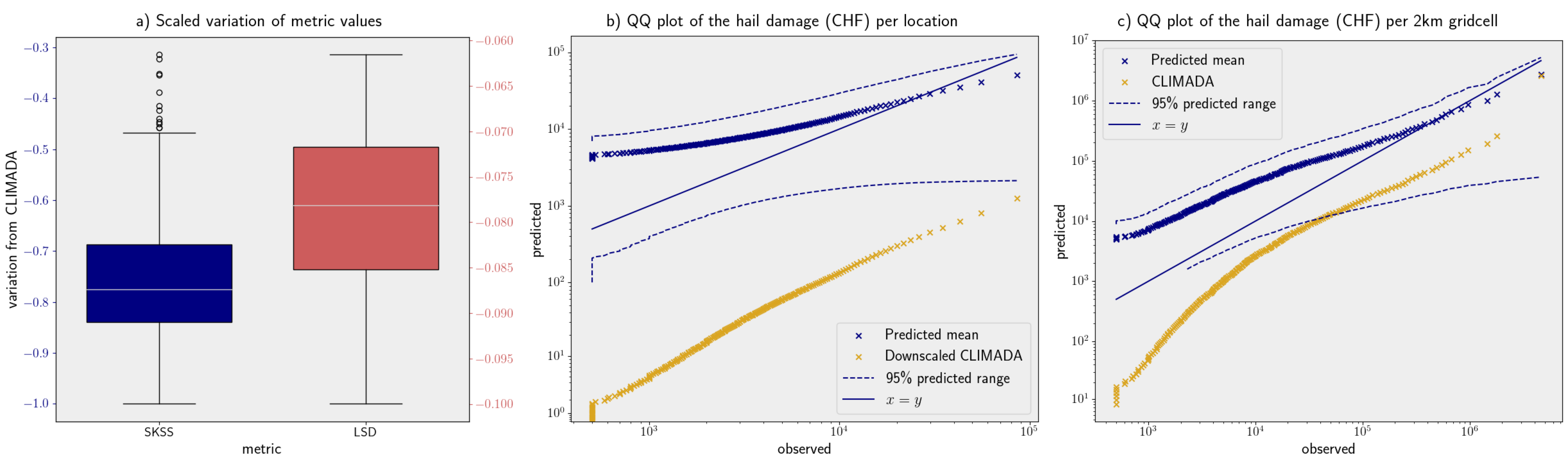}
    \caption{Comparison of metrics of predicted damage.  (a) values of scaled LSD and SKSS for our model, with those from CLIMADA subtracted. (b) QQ-plots of realized versus predicted quantiles for the damage per building and (c) per 2km grid cell, with dashed blue lines showing the 95\% prediction range.}
    \label{fig:metrics}
    \end{figure}

Evaluation of the full hail damage prediction involves the combination of counts and claim values, as described in Section~\ref{sec:combination}. We use the two metrics described in Section~\ref{sec:metrics} to compare the spatial patterns of hail damage predicted with CLIMADA and our model. Figure~\ref{fig:metrics}(a) shows that both take higher values for CLIMADA than for our model, so the latter better captures the observed hail damage footprints. 
Figure~\ref{fig:metrics} also shows that the distribution of hail damage predicted with our model is close to the observed distribution both on a single building scale (b) and on a 2km square grid (c), though non-extreme damage is slightly over-predicted, which suggests further research on the distribution of non-extreme claim values might be needed. CLIMADA systematically under-predicts the aggregated damage per grid cell, and there is a clear improvement using the random line process. Figure~\ref{fig:metrics}(b) compares the model's input, downscaled CLIMADA, to the prediction. As expected, downscaled CLIMADA under-predicts damage per building (see Section~\ref{sec:extremes}), while the random line model predicts realistic values, particularly so for claims above CHF~5000.

\begin{center}
    \begin{figure}[!th]
        \centering
        \includegraphics[width= \linewidth]{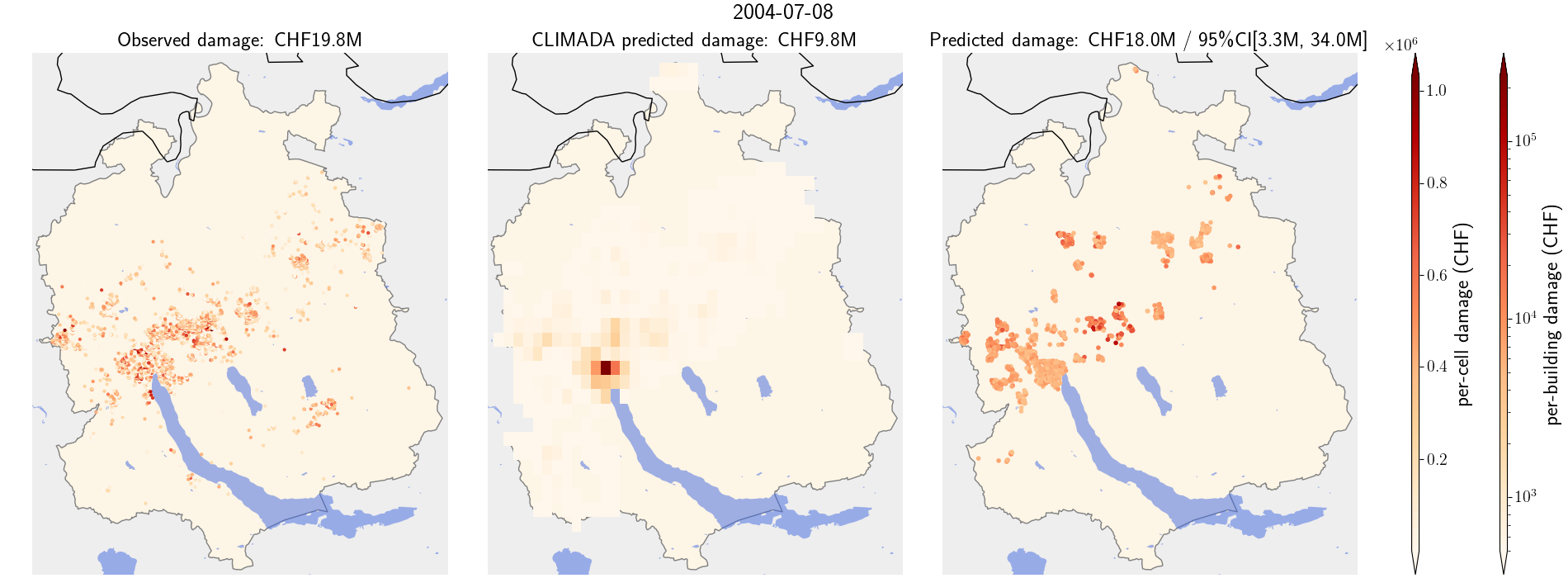}
        \includegraphics[width= \linewidth]{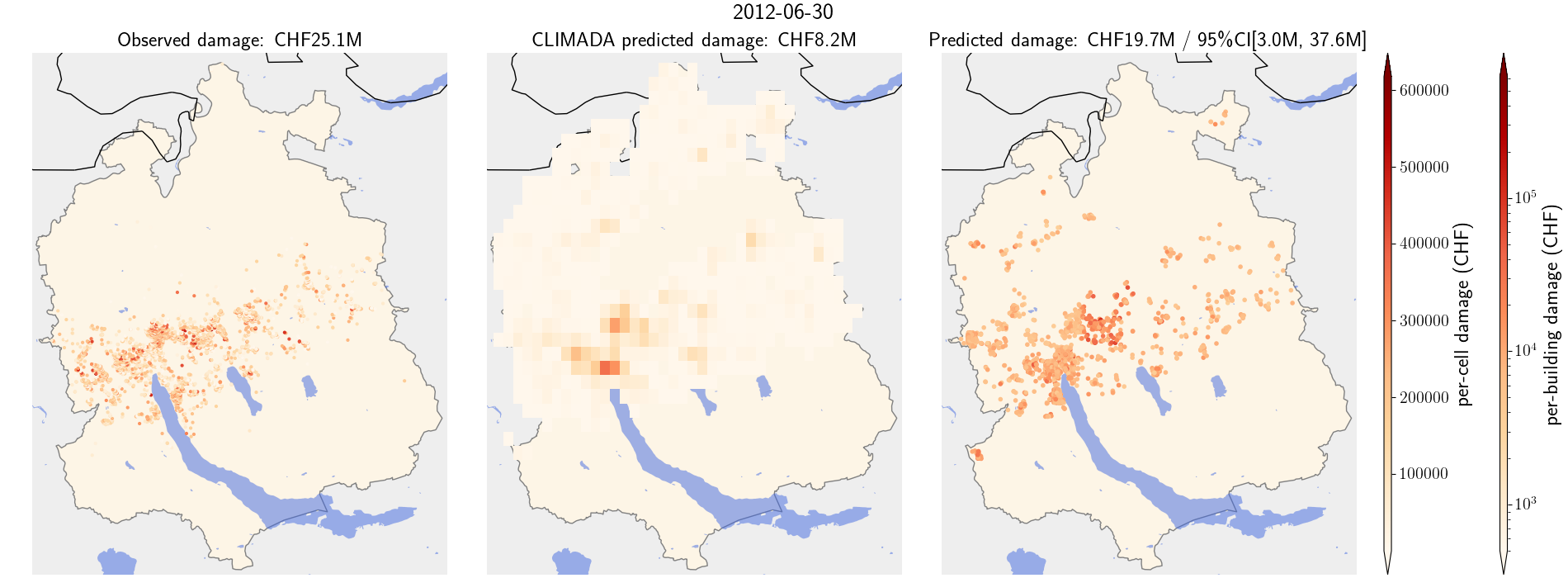}
        \includegraphics[width= \linewidth]{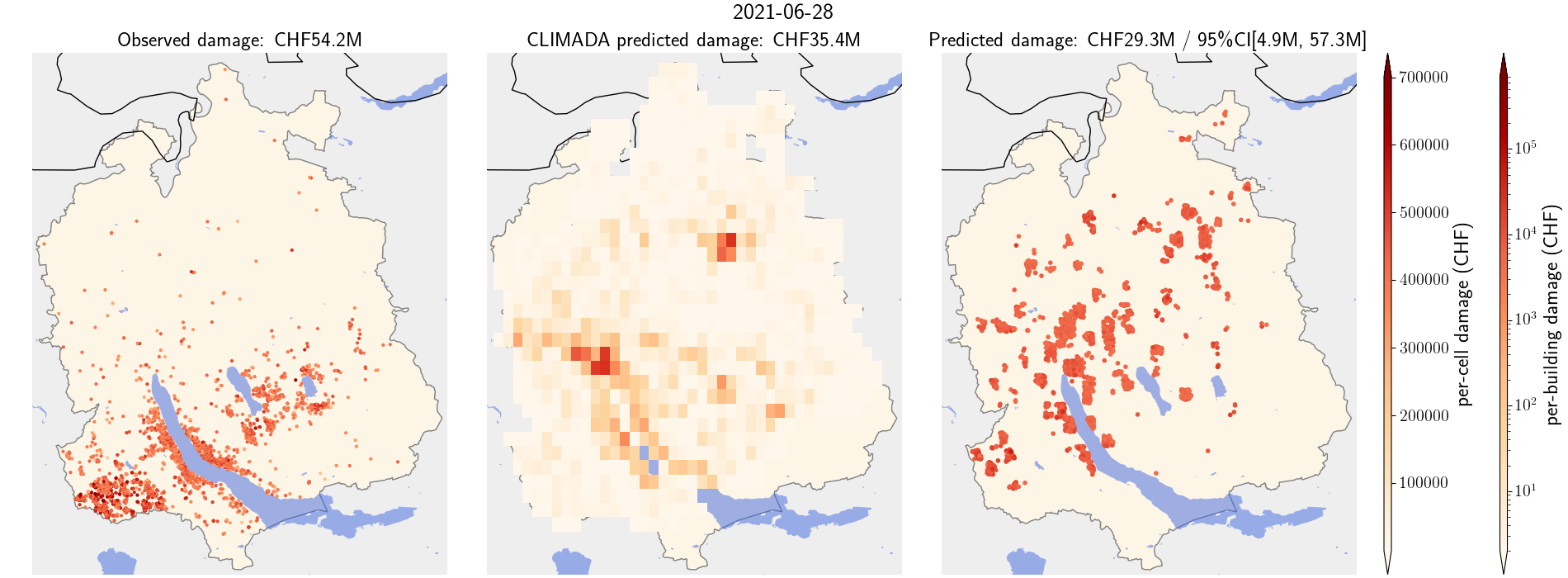}
        \caption{Example daily hail impact maps. The columns represent the observed claims (left), CLIMADA-predicted claims (center), and the hail damage prediction using the random line model (right). The left color bar relates to the prediction at the scale of the cell (i.e. relevant for CLIMADA predicted damage), while the right color bar displays a log scale for the per-building claim values (i.e. relevant for realized and predicted damage). The observed monetary cantonal damage (left), CLIMADA predicted total (middle), and average damage from the random line model with its 95\% predicted range (right) are displayed in the titles for each selected date.}
        \label{example_maps} 
    \end{figure}
\end{center}

Figure~\ref{example_maps} shows some daily hail damage maps for days on which there was over one million Swiss francs of realized hail impacts in the canton (some of these claim dates belong to the train or validation set). The predicted claim values appear to be locally large, matching the spatial pattern of realized damage, whereas CLIMADA damage is more dispersed. The average total predicted damage over the canton for days with extreme realized hail damage is close to the observed value, which the confidence interval usually captures. The line model is thus able to predict well-located extremes while providing reliable estimates of the total damage. 
The lowest panel in Figure~\ref{example_maps} shows the most extreme hail event in the two decades of our data, on 28 June 2021, which involved roof-penetrating hail damage with 177 claims above CHF~100,000. Our model manages to capture extreme damage on this day, with average predicted values up to CHF~180,000. The spatial pattern using both CLIMADA and, to a lesser extent, our model, is wider than the observed data, which might be related to overestimation of the MESHS intensity that day in the northern half of the canton (Figure~\ref{fig:exploratory_climate_variables}).

\section*{Conclusion}
The model developed in this paper seems to be the first to combine a Gaussian random line process with extreme-value modeling in order to predict the spatial footprint of hail damage.  It improves on the use of a benchmark deterministic hail damage function: in particular, it captures extreme damage values for individual buildings well, reproduces the spatial pattern of hail in the insurance data, and its stochasticity enables uncertainty quantification. With appropriate changes, such as for instance the possibility of modeling multiple random lines at the same time, our approach could be generalized to larger areas and would be useful in studying the insurance impacts of climate change. It would be interesting to use thunderstorm cell direction instead of large-scale wind direction as the covariate for the slope of the random line process. 

\section*{Authorship contribution statement}
\paragraph*{Oph{\'e}lia Miralles} Methodology, Software, Validation, Formal Analysis, Investigation, Data Curation, Writing (original draft, review, editing), Visualization.
\paragraph*{Anthony Davison} Methodology, Supervision, Editing, Funding acquisition.
\paragraph*{Timo Schmid} Data Curation, Editing.

\begin{acks}[Acknowledgments]
We thank the Federal Office of Meteorology and Climatology (MeteoSwiss) for providing the POH and MESHS data, the insurance company GVZ for providing the building damage and exposure data, and Daniel Steinfeld of GVZ for valuable insights.
\end{acks}

\begin{funding}
Oph{\'e}lia Miralles' work was partially funded by the Swiss National Science Foundation (project 200021\_178824). Timo Schmid acknowledges funding from the Swiss National Science Foundation (SNSF) Sinergia grant (CRSII5\_201792).
\end{funding}
 
\section*{Data statement}
 The exploratory analysis, model, and diagnostics are implemented in Python and the code is freely available on GitHub (\href{https://github.com/OpheliaMiralles/hail-damage-modeling/}{github.com/OpheliaMiralles/hail-damage-modeling}). MESHS and POH data are available from MeteoSwiss on demand. The GVZ insurance data are private, and as such are available for use only within the scClim project for research purposes. The CLIMADA impact function is open-source and can be run using the GitHub repository \href{https://github.com/CLIMADA-project/climada_papers}{github.com/CLIMADA-project/climada{\_}papers}. ERA5 reanalysis data can be downloaded freely from the Copernicus Climate Data Store (\href{https://climate.copernicus.eu/climate-reanalysis}{climate.copernicus.eu/climate-reanalysis}).
 
\clearpage
\bibliographystyle{imsart-nameyear}
\bibliography{bibli}

\clearpage
\begin{supplement}
\stitle{Additional details}
\sdescription{Here we provide additional information and figures about model selection, including material about the choice of fixed hyperparameters such as the threshold for the GPD model, and also MC diagnostics related to the validation of the Bayesian model parameters.}
\end{supplement}

\begin{figure}[!ht]
    \centering
\includegraphics[width=\linewidth]{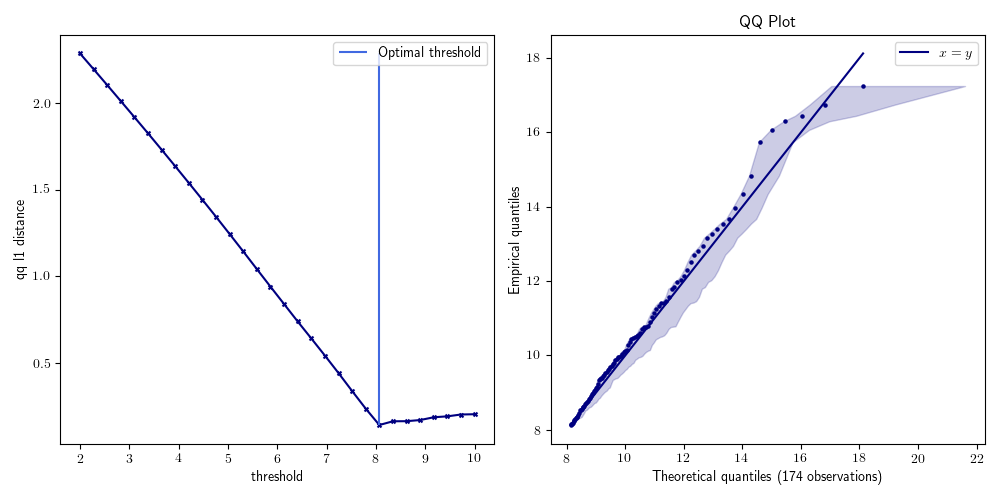}
    \caption{Threshold selection method described in \citet{varty2021inference} applied to the log total damage per cell. The left panel shows a minimum qq-$\ell_1$-distance for a threshold of 8.06. The QQ-plot for the GPD fit of the log total sum of damage above this threshold is displayed in the right panel, in which the profile likelihood-based 95\% confidence interval is shown by the blue shaded area.}
    \label{fig:threshold_sel}
\end{figure}

\begin{figure}[!ht]
    \centering
\includegraphics[width=\linewidth]{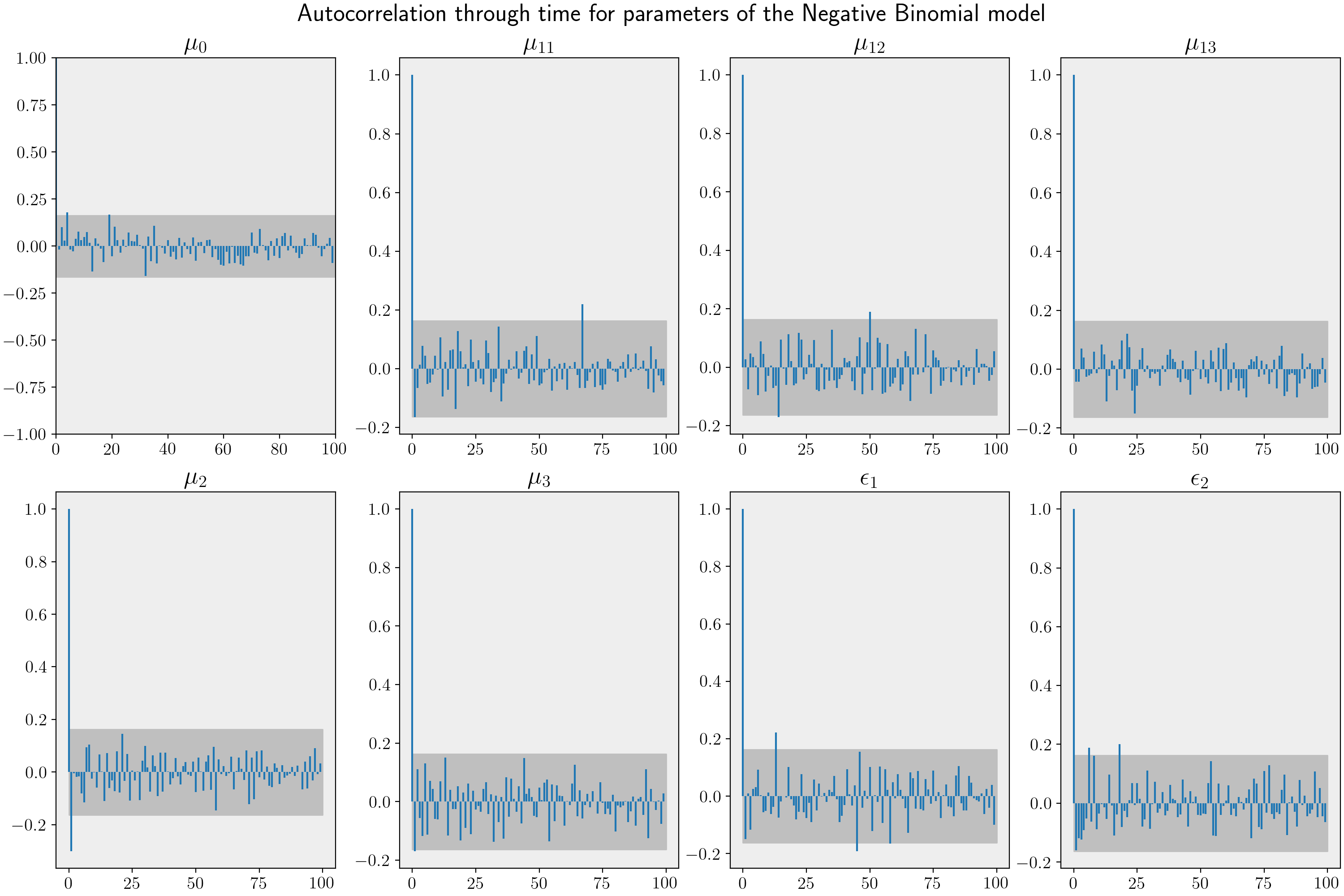}
    \caption{Evolution of the autocorrelation through sampling for parameters of the Negative Binomial model presented in Section~\ref{sec:count_model}. The grey area designates the acceptable range for the autocorrelation at the end of sampling to assume convergence of the model. The bounds of the confidence range are computed from the central limit theorem.}
\end{figure}

\begin{figure}[!ht]
    \centering
\includegraphics[width=\linewidth]{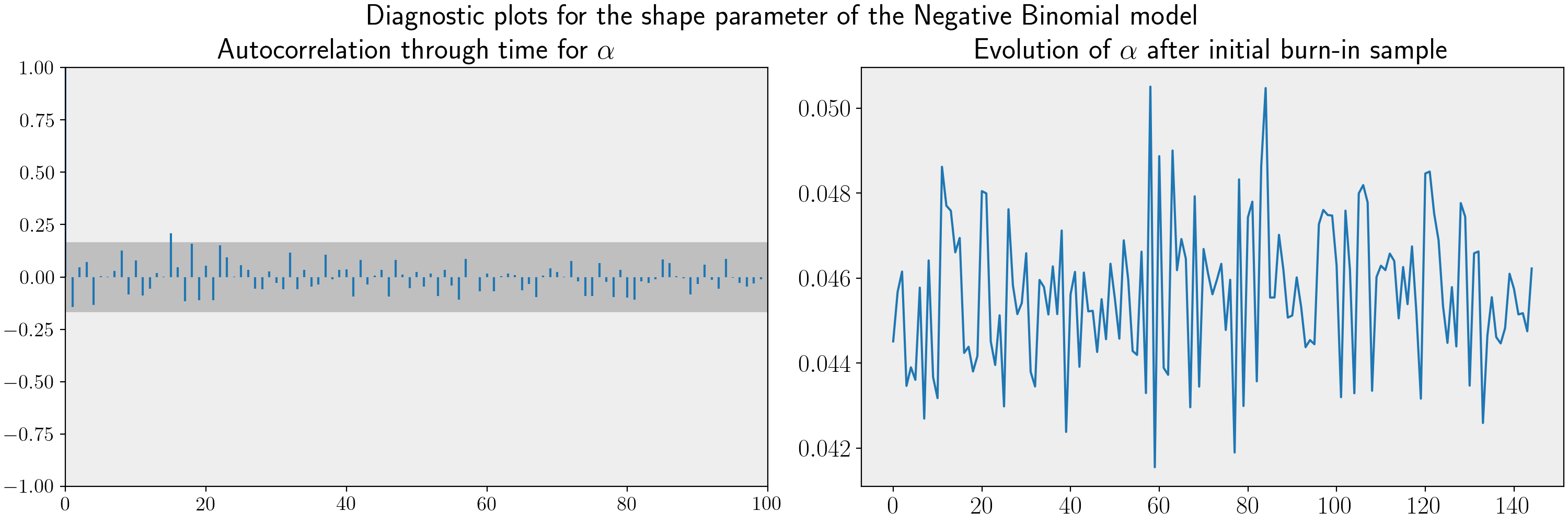}
    \caption{Autocorrelation plot (left) and trace plot (right) for the posterior distribution of the shape parameter $\alpha$ in the Negative Binomial model presented in Section~\ref{sec:count_model}. A close-to-zero autocorrelation through sampling and no specific trend or pattern in the trace plot is usually a good sign of model convergence.}
\end{figure}

\begin{figure}[!ht]
    \centering
\includegraphics[width=\linewidth]{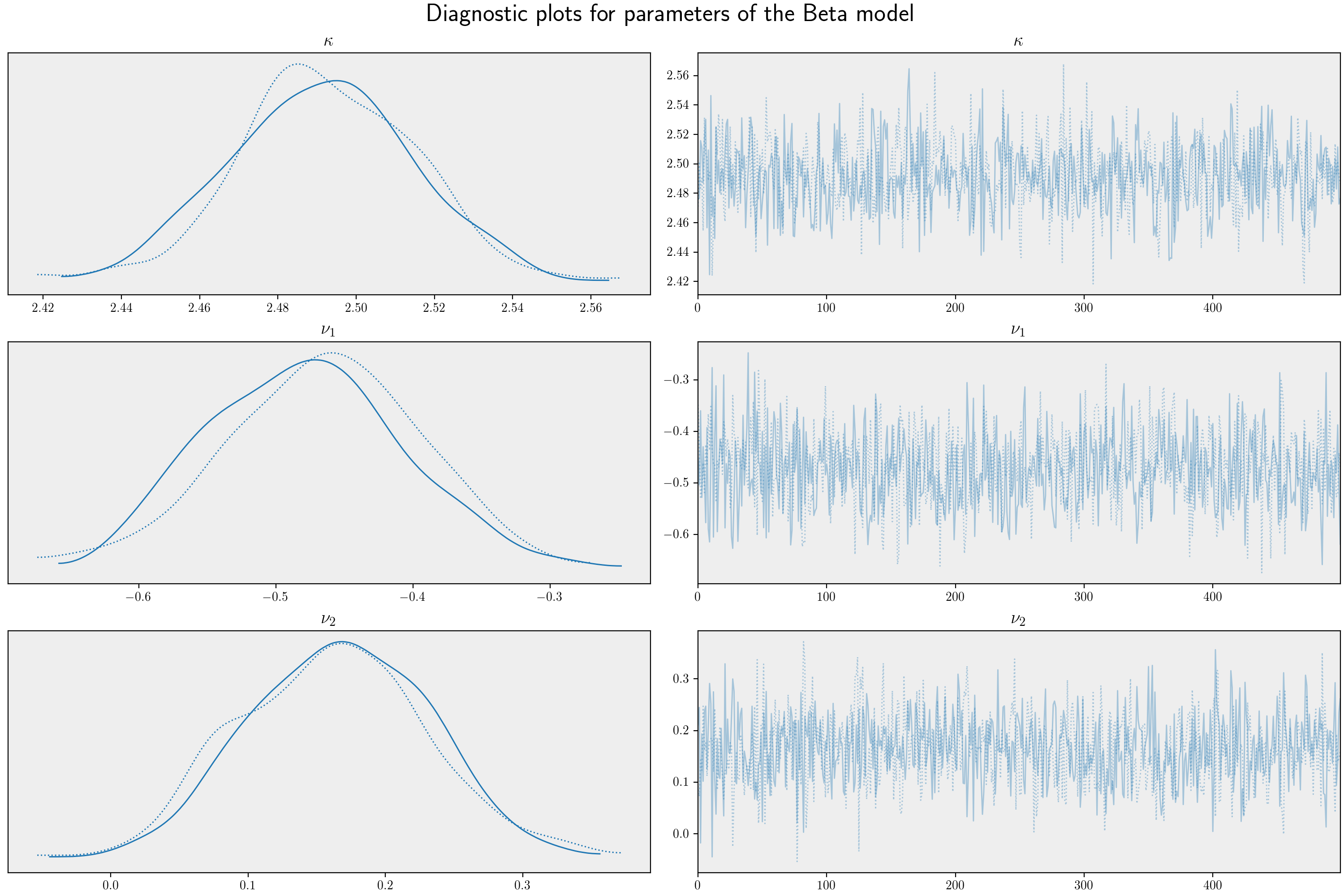}
    \caption{Kernel density estimate plot: (left) and trace plot: (right) for parameters of the Beta model presented in Section~\ref{sec:beta_model}. No specific trend or pattern in the trace plot is usually a good sign of model convergence.}
\end{figure}

\begin{figure}[!ht]
    \centering
\includegraphics[width=\linewidth]{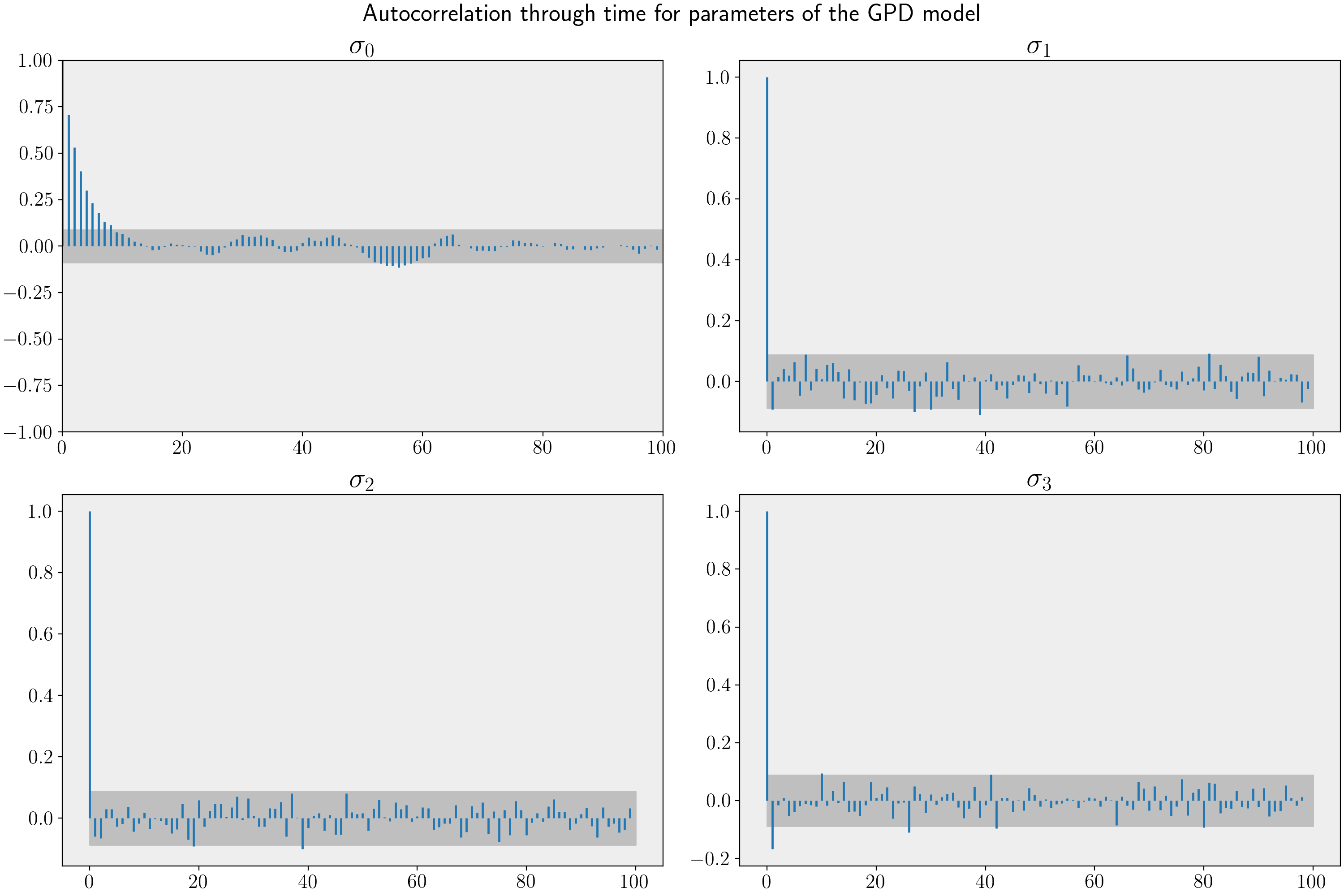}
    \caption{Evolution of the autocorrelation through sampling for parameters of the GPD model presented in Section~\ref{sec:gpd_model}. The grey area designates the acceptable range for the autocorrelation at the end of sampling to assume convergence of the model. The bounds of the confidence range are computed from the central limit theorem.}
\end{figure}

\begin{figure}[!ht]
    \centering
\includegraphics[width=\linewidth]{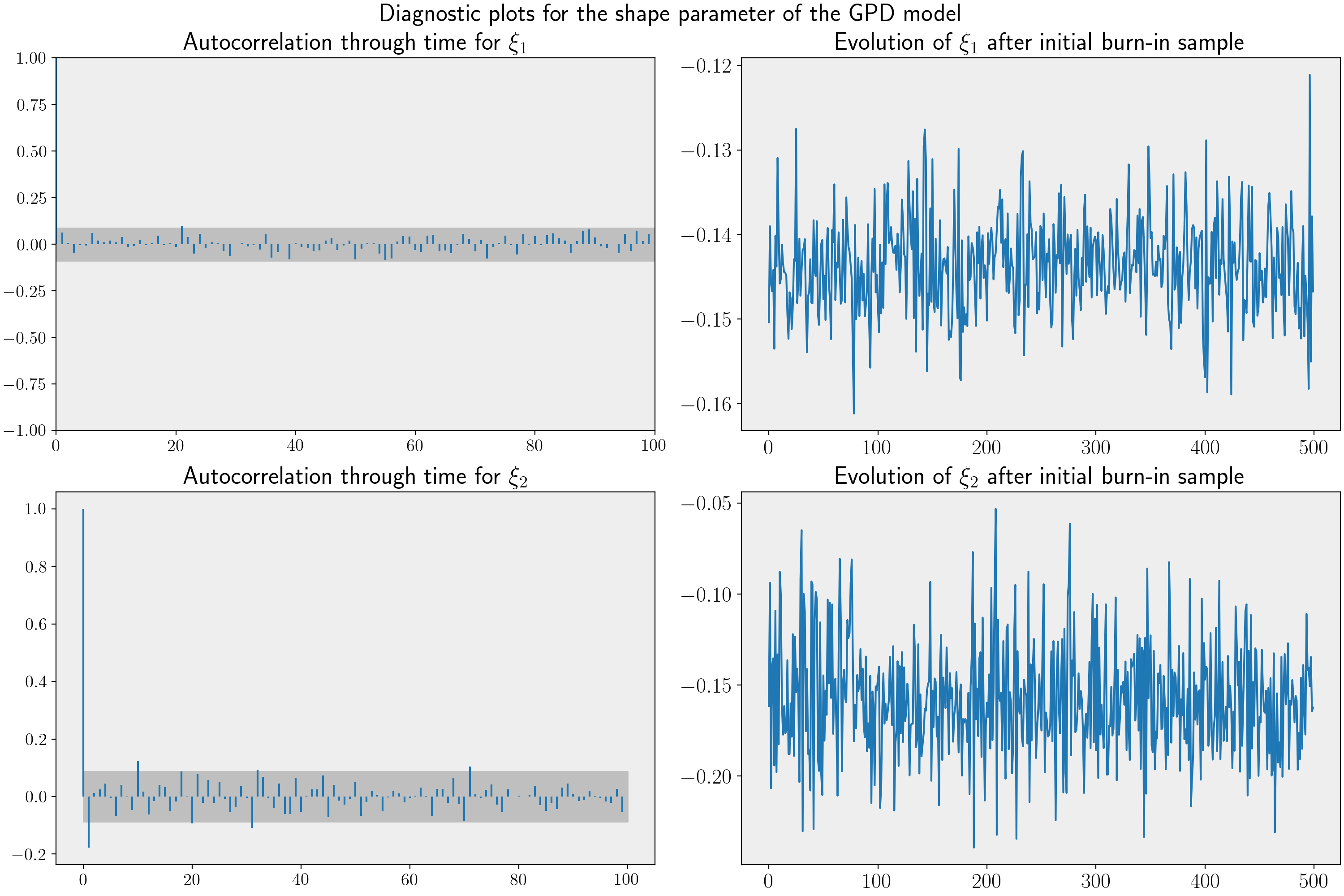}
    \caption{Autocorrelation plot (left) and trace plot (right) for the posterior distribution of the shape parameter $\xi$ in the GPD model presented in Section~\ref{sec:gpd_model}. A close-to-zero autocorrelation through sampling and no specific trend or pattern in the trace plot is usually a good sign of model convergence.}
\end{figure}

\end{document}